\documentclass[a4paper,11pt,oneside,onecolumn,number,preprint]{elsarticle}
\usepackage[margin=1in]{geometry}
\usepackage{setspace}
\usepackage{amsmath}
\usepackage{textcomp}
\usepackage{amstext}
\usepackage{amssymb}
\usepackage{graphicx}
\usepackage{hyperref}

\journal{Journal of \LaTeX\ Templates}

\begin{document}

\begin{frontmatter}

\title{Secrecy Outage of SWIPT in the Presence of Cooperating Eavesdroppers}


\author[1]{Furqan Jameel}

\author[1]{Shurjeel Wyne\corref{correspondingauthor}}
\ead{shurjeel.wyne@comsats.edu.pk}

\address[1]{Dept. of Electrical Engineering, COMSATS Institute of Information Technology, Islamabad, Pakistan}
\cortext[correspondingauthor]{Corresponding author.}

\begin{abstract}
This work investigates the secrecy performance for a simultaneous wireless information and power transfer system that operates in the presence of cooperating eavesdroppers. The multi-antenna access point's transmission is used for information-decoding by a multi-antenna node and for energy-harvesting (EH) by multiple single-antenna nodes. However, some of the nodes authorized for EH only attempt to eavesdrop on the ongoing information transmission by utilizing a generalized on-off power splitting architecture. We derive a closed-form expression for the secrecy outage probability of the considered Multiple-Input-Multiple-Output Multiple-Eavesdroppers system. Theoretical and simulation results are provided to validate the derived results.
\end{abstract}


\end{frontmatter}


\section{Introduction}

Recently, the development of smart receivers to harvest energy from radio frequency (RF) signals has drawn significant research interest. The dual role of RF signals to transfer information and energy has led to the emergence of the concept of simultaneous wireless information and power transfer (SWIPT). The tradeoff between achievable rate and harvested energy for a SWIPT system with Amplify-and-Forward (AF) relays was evaluated in \cite{nguyen2016two}; Nguyen et al. proposed two protocols based on a power-splitting technique for delay tolerant and delay limited networks. The outage performance of dual-hop cooperative networks was also investigated by \cite{pan2017outage}. However, ensuring link security in SWIPT systems is a challenging problem \cite{chen2016secrecy} and exploiting the wireless channel characteristics has emerged as an effective means to attain message confidentiality in so called Physical Layer Security systems \cite{chen2016secrecy}. In \cite{dai2015feedback}, Dai et al. consider the classical Wyner's wiretap channel model and investigate the significance of using feedback of channel state information (CSI) from the receiver to enhance message secrecy. Network coding was studied in \cite{cao2015secure} to secure the messages from an eavesdropper, without reducing the throughput. Secrecy performance of Decode and Forward (DF) relaying under Rayleigh fading was investigated in \cite{zhang2017secrecy} and a best-node selection scheme was proposed. The achievable secrecy rate for SWIPT was studied in \cite{liu2013secrecy} for Multiple-Input-Single-Output (MISO) systems in the presence of multiple eavesdroppers. Zhang et al. in \cite{zhang2016energy} studied the achievable secrecy rate of an energy-harvesting (EH) orthogonal frequency division multiplexing system. All sub-carriers were allocated the same power with power splitting (PS) utilized to coordinate information decoding (ID) and EH. 

The aforementioned studies are limited in that they assume the eavesdroppers to operate independently. However, in practical scenarios the multiple eavesdroppers, if present, may cooperate with one another to decode the message \cite{chen2016secrecy}. Despite this, SWIPT performance in the presence of multiple cooperating, i.e., information combining and exchanging, eavesdroppers has only recently been investigated \cite{7393611,khandaker2015masked}; those works consider eavesdroppers that apply joint signal processing techniques to decode the confidential message. Such type of signal processing requires the eavesdroppers to be equipped with sophisticated hardware and accurate time synchronization, which may not be possible with energy limited wireless devices. The novel contribution of this work is the derivation of a closed-form expression for the secrecy outage probability of a transmit-antenna-selection (TAS) SWIPT system that operates in the presence of cooperating eavesdroppers, which employ on-off power splitting (OPS). Our work differs from \cite{7393611,khandaker2015masked} in that we consider cooperative scheduling among the eavesdroppers such that the eavesdropper with highest instantaneous signal-to-noise ratio (SNR) is scheduled to decode the confidential message. Also, in contrast to \cite{7393611,khandaker2015masked}, our analysis treats the generic $\kappa-\mu$ fading model on the links \cite{milisic2008outage}; this fading model contains other conventionally used channel models as special cases. For example, the $\kappa-\mu$ model resolves to Rayleigh fading for $\mu=1$ and $\kappa\rightarrow 0$, the Rician fading model is obtained for $\mu=1$ and $\kappa>0$, whereas the Nakagami-$m$ fading model is obtained for $\mu>0$ and $\kappa\rightarrow 0$ \cite{milisic2008outage}. Furthermore, while most published works utilize either time switching (TS) \cite{zhou2013wireless} or PS \cite{7393611,zhou2013wireless} to incorporate EH and ID simultaneously at the receiver (RX), our analysis is based on the more generalized OPS architecture, which contains TS and PS as special cases \cite{zhou2013wireless}.
\section{System Model}
\begin{figure}
\centering
\includegraphics[width=.4\textwidth]{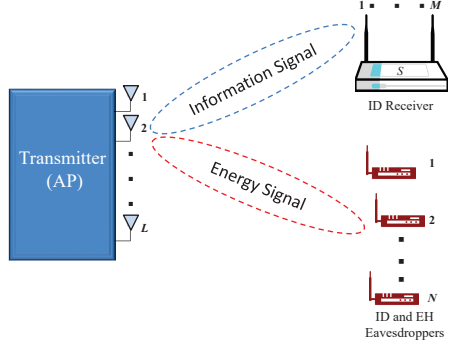}
\caption{System Model.}
\label{fig:fig1}
\end{figure}

We consider the downlink of a SWIPT system comprising an $L$-antenna access point (AP) transmitting to an ID-only $M$-antenna node $S$ and multiple ID and EH single-antenna nodes as shown in Fig. \ref{fig:fig1}. Among the nodes admitted into the network for EH only, a set of $N$ nodes, $E=\left\{E_i|i=1,...N\right\}$, attempt to cooperatively eavesdrop on the ongoing information transmission. During a single scheduling slot, the AP selects one out of $L$ antennas to convey information to $S$ and transfer energy to $E$. The TAS scheme at AP is used to maximize achievable secrecy rate on the main link between AP and $S$; it leverages spatial selectivity of the channel through $L$ diversity paths while simultaneously reducing hardware costs by employing a single RF chain at AP. The AP is assumed to have CSI for main channel to $S$ as well as CSI of wiretap channels from AP to the eavesdroppers \cite{7393611,khandaker2015masked}. The $N$ eavesdroppers cooperate among themselves to exchange and combine information for decoding the intercepted message. For this purpose the eavesdroppers are assumed to be equipped with an OPS architecture that allows them to simultaneously incorporate EH and ID reception \cite{zhou2013wireless}. According to the OPS scheme, for the first $\alpha K$ symbols transmitted in downlink, $0\leq\alpha\leq1$, the RF power is used only for EH and the ID receiver remains off during this period. For the remaining $\left(1-\alpha\right)K$ symbols, the RF signal is power-split into two streams where one stream with power ratio $0\leq\rho\leq1$ is used for ID and the other stream with power ratio $\left(1-\rho\right)$ is used for EH. With this arrangement, the OPS scheme contains the TS and PS schemes as special cases for $\rho=1$ and $\alpha=0$, respectively. Let the AP transmits to $S$ with average power $P$ using its $l$-th antenna; then the signal received at $S$ with fading and pathloss effects included is written as%
\begin{equation}%
  \mathbf{y}^l_s=\sqrt{\frac{P}{{PL}_s}}\mathbf{h}^l_sx+\mathbf{n}_s,%
\end{equation}%
where $\mathbf{h}^l_s=\left[h^l_{s_1},h^l_{s_2},\ldots,h^l_{s_M}\right]^T$ represents the vector channel between the AP and $S$. Furthermore, $\mathbf{n}_s=\left[n_{s_1},n_{s_2},\ldots,n_{s_M}\right]^T$ represents the zero-mean additive white Gaussian noise (AWGN) with variance $N_0$ at $S$ due to the receiver electronics. Furthermore, ${PL}_s=\left(\frac{4\pi}{G_t G_r\lambda}\right)^2d^\beta$ is the distance-dependent path loss of the main link \cite{chen2016secrecy}, where $d$ is the distance between transmitter (TX) and RX and $\beta$ is the pathloss exponent, $G_t$ and $G_r$ are TX and RX antenna gains and $\lambda$ is the carrier wavelength. Since the AP transmission is broadcast, the signal received at $i$-th eavesdropper is written as \cite{zhou2013wireless}%
\begin{align}%
y^l_{ie}=\sqrt{{\rho }_i}\left(\sqrt{\frac{P}{{PL}_{ie}}}h^l_{ie}x+n_{ie}\right)+z_i,%
\end{align}%
where ${PL}_{ie}$ is the pathloss and $h^l_{ie}$ represents the fading channel coefficient between $l$-th antenna on AP and $i$-th eavesdropper. Furthermore, $n_{ie}$ represents the zero-mean AWGN with variance $N_0$ at $i$-th eavesdropper and $z_i$ is the signal processing noise at $i$-th eavesdropper due to its decoder's rectifier, which is also modeled as AWGN with zero-mean and variance $\sigma_i^2=\sigma^2$ for all eavesdroppers since they are assumed to use the same type of rectification circuit. Finally, for tractable analysis we consider $\rho_i=\rho$ and ${PL}_{ie}=PL_e,\,\forall i\in N$. The node $S$ employs maximum ratio combining (MRC) of the independent $\kappa-\mu$ fading $M$ branches such that the signal at combiner output is written as $\hat{x}=\frac{{\mathbf{h}^l_s}^H\mathbf{y}^l_s}{\left\|\mathbf{h}^l_s\right\|}$ with corresponding instantaneous SNR given by $\gamma^l_s=\frac{\sum^M_{j=1}{\left|h^l_{sj}\right|}^2P}{PL_sN_0}$. Then $\gamma^l_s$ has a $\kappa-\mu$ probability density function (PDF) written as \cite{milisic2008outage}%
\begin{align}%
f_{\gamma_{s}^{l}}(\gamma _{s}^{l})=\frac{M\mu _{s}(1+\kappa_{s})^{\frac{M\mu _{s}+1}{2}}(\gamma _{s}^{l})^{\left(\frac{M\mu_{s}-1}{2}\right)}}{\kappa_{s}^{\left(\frac{M\mu_{s}-1}{2}\right)}\mathrm{e}^{(M\mu _{s}\kappa _{s})}\left(M\frac{\bar{\gamma}_{s}^{l}}{PL_s}\right)^{\left(\frac{M\mu _{s}+1}{2}\right)}} \mathrm{e}^{\left(-\frac{\mu _{s}(1+\kappa_{s})\gamma _{s}^{l}}{\frac{\bar{\gamma}_{s}^{l}}{PL_s}}\right)}I_{\mu _{s}-1}\left[ 2\mu_{s}\sqrt{\frac{M\kappa_{s}(1+\kappa _{s})\gamma _{s}^{l}}{\frac{\bar{\gamma }_{s}^{l}}{PL_s}}}\right], \label{eq_f_gamma_l_s_def}%
\end{align}%
where $\overline{\gamma}^l_s$ represents the average SNR at combiner output and $I_v(.)$ is the modified Bessel function of the first kind \cite[Eq.(8.406)]{gradshteyn2014table}. It may be mentioned here that in the $\kappa-\mu$ fading model, $\mu$ represents the number of multipath clusters and $\kappa$ represents the ratio between the total power of dominant component and scattered waves in each cluster \cite{milisic2008outage}. 

Similarly the instantaneous SNR at the $i$-th eavesdropper given as $\gamma^l_{ie}=\frac{\rho\left|h^l_{ie}\right|^2P}{PL_e N_0\left(\rho +\frac{\sigma^2}{N_0}\right)}$ has a $\kappa-\mu$ distributed PDF given as \cite{milisic2008outage}%
\begin{align}%
f_{\gamma_{ie}^{l}}(\gamma_{ie}^{l})=\frac{\mu_{ie}(1+\kappa_{ie})^{\frac{\mu_{ie}+1}{2}}(\gamma_{ie}^{l})^{\left(\frac{\mu_{ie}-1}{2}\right)}}{\kappa_{ie}^{\left(\frac{\mu_{ie}-1}{2}\right)}\mathrm{e}^{\left(\mu_{ie}\kappa_{ie}\right)}(\omega \bar{\gamma}_{ie}^{l})^{\left(\frac{\mu_{ie}+1}{2}\right)}} \mathrm{e}^{\left(-\frac{\mu_{ie}(1+\kappa _{ie})\gamma _{ie}^{l}}{\omega \bar{\gamma}_{ie}^{l}}\right)} I_{\mu_{ie}-1}\left[2\mu_{ie}\sqrt{\frac{\kappa_{ie}(1+\kappa_{ie})\gamma_{ie}^{l}}{\omega\bar{\gamma}_{ie}^{l}}}\right]%
\end{align}%
where $\omega = \frac{\rho}{PL_e\left(\rho+\frac{\sigma^2}{N_0}\right)}$. In each scheduling slot the cooperating eavesdroppers share their instantaneous SNR values so that the node with the highest instantaneous SNR is chosen to decode the source's message. In this case, the achievable rate for the wiretap link is determined by the maximum instantaneous SNR among $N$ nodes, which can be written as $C^l_e=\left(1-\alpha\right)\textrm{log}_2\left(1+\textrm{max}_{i\in N}\gamma^l_{ie}\right)$. Expressing the main link's achievable rate as $C^l_s=\textrm{log}_2\left(1+\gamma^l_s\right)$, the achievable secrecy rate is defined as the non-negative difference between these two rates and is expressed as $C_{sec,l}=max\left[C^l_s-C^l_e,0\right]$.
\section{Outage Probability Analysis}%
The outage event occurs when $C_{sec}\triangleq\mathop{\mathrm{max}}_{l=1,2,\dots,L}C_{sec,l}$ falls below some target rate $R_s>0$. The probability of this event is given as
\begin{align}%
P_{out}=\mathrm{Pr}\left[C_{sec}<R_s\right]=\mathrm{Pr}\left[\mathop{\mathrm{max}}_{l=1,2,\dots,L}C_{sec,l}<R_s\right]=\prod^L_{l=1}{P_{out,l}},\label{eq_P_out_def}%
\end{align}%
where $\mathop{\mathrm{max}}_{l=1,2,\dots,L}C_{sec,l}$ is due to TAS at the AP, $P_{out,l}\triangleq\mathrm{Pr}\left[C_{sec,l}<R_s\right]$, and i.i.d. fading on $L$ branches is used to obtain (\ref{eq_P_out_def}). Let coverage probability at $l$-th antenna, $P_{cov,l}\triangleq 1-P_{out,l}$ is written as%
\begin{align}%
P_{cov,l}=\mathrm{Pr}\left[\mathrm{log}_2\left(\frac{1+\gamma^l_s}{\left(1+\mathop{\mathrm{max}}_{i\in N}\gamma^l_{ie}\right)^{\left(1-\alpha\right)}}\right)>R_s\right]=\prod^N_{i=1}P^l_{eve,i},\label{eq_P_cov_l_def}%
\end{align}%
where $P^l_{eve,i}=\mathrm{Pr}\left[\gamma^l_{ie}<\left(\frac{1+\gamma^l_s}{2^{R_s}}\right)^{\frac{1}{1-\alpha}}-1\right]=\int^{\infty }_{2^{R_s}-1}{\int^{\gamma_{th}}_0{f_{\gamma^l_s,\gamma^l_{ie}}\left({\gamma }^l_s,{\gamma }^l_{ie}\right)}}\ d{\gamma }^l_{ie}d{\gamma }^l_s$ and $\gamma_{th}={\left(\frac{1+\gamma^l_s}{2^{R_s}}\right)}^{\frac{1}{1-\alpha }}-1$. Now by exploiting the independence of $\gamma^l_s,\gamma^l_{ie}$ we get  
\begin{align}%
P^l_{eve,i}=\int^\infty_{2^{R_s}-1}{F_{\gamma^l_{ie}}\left[{\left(\frac{1+\gamma^l_s}{2^{R_s}}\right)}^{\frac{1}{1-\alpha}}-1\right]}f_{\gamma^l_s}\left(\gamma^l_s\right)d\gamma^l_s,\label{eq_P_l_eve_i_def}%
\end{align}%
where $F_{\gamma^l_{ie}}\left[\gamma\right]\triangleq 1-Q_{\mu_{ie}}\left[\sqrt{2\kappa_{ie}\mu_{ie}}, \sqrt{\frac{2(\kappa_{ie}+1)\mu_{ie}\gamma}{\bar{\gamma}_{ie}^{l}}}\right]$ is the cumulative distribution function of the $\kappa-\mu$ distributed $\gamma^l_{ie}$ \cite{milisic2008outage} and $Q_v(a,b)$ is the well-known Marcum-$Q$ function. Now plugging $f_{\gamma^l_s}\left(\gamma^l_s\right)$ from (\ref{eq_f_gamma_l_s_def}) into (\ref{eq_P_l_eve_i_def}) and using \cite[Eq.(3.381)]{gradshteyn2014table} we obtain after some manipulations the expression
\begin{align}%
P^l_{eve,i}&=\frac{M}{\mathrm{e}^{(\mu _{ie}\kappa _{ie}+M\mu _{s}\kappa_{s})}} \times \sum_{t=0}^{\infty }{\frac{(\mu _{ie}\kappa _{ie})^{t}}{t!\Gamma(\mu_{ie}+t)}}\times \sum_{v=0}^{\infty }{\frac{(\mu _{s})^{2v+\mu_{s}}}{v!\Gamma(\mu_{s}+v)}}\times \frac{(\Psi _{s})^{\Phi _{2}}}{\kappa_{s}^{\Phi_{1}}M^{\Phi_{2}}} \nonumber \\
&\times \int_{2^{R_{s}}-1}^{\infty}{(\gamma _{s}^{l})^{\Phi _{1}}}(M\kappa _{s}\Psi _{s}\gamma_{s}^{l})^{\left(\frac{2v+\mu _{s}-1}{2}\right)}\times \gamma(\mu_{ie}+t,\Psi_{e}\mu_{e}\gamma_{th})\mathrm{e}^{(-\mu_{s}\Psi_{s}\gamma _{s}^{l})} d\gamma_{s}^{l}. \label{eq_P_l_eve_i_def3}%
\end{align}%
In (\ref{eq_P_l_eve_i_def3}), $\gamma\left(.,.\right)$ is the incomplete Gamma function and $\Gamma\left(.\right)$ is the Gamma function; furthermore, $\Psi _{s}=\frac{1+\kappa _{s}}{\bar{\gamma }_{s}^{l}/PL_s}, \Psi _{e}=\frac{1+\kappa _{e}}{\omega \bar{\gamma }_{ie}^{l}}, \Phi _{1}=\frac{M\mu _{s}-1}{2}$ and $\Phi _{2}=\frac{M\mu _{s}+1}{2}$. Finally, using (\ref{eq_P_l_eve_i_def3}) and (\ref{eq_P_cov_l_def}) in (\ref{eq_P_out_def}) the outage probability at $S$ with TAS employed at the AP is obtained as
\begin{align}%
P_{out}=\prod^L_{l=1}{\left[1-\prod^N_{i=1}{P^l_{eve,i}}\right]}.%
\end{align}%
Apart from the secrecy outage probability, another metric for evaluating secrecy performance is the secrecy throughput defined as the minimum rate guaranteed to be supported by all legitimate users \cite{liu2014secrecy}. Recalling the classical Wyner model for the case of non-adaptive on-off transmission, i.e., $R_s$ remains constant during the transmission and is independent of $\gamma_{s}^{l}$, the secrecy throughput can then be written as \cite{zheng2014secrecy}
\begin{figure}
	\centering
		\includegraphics[width=.5\textwidth]{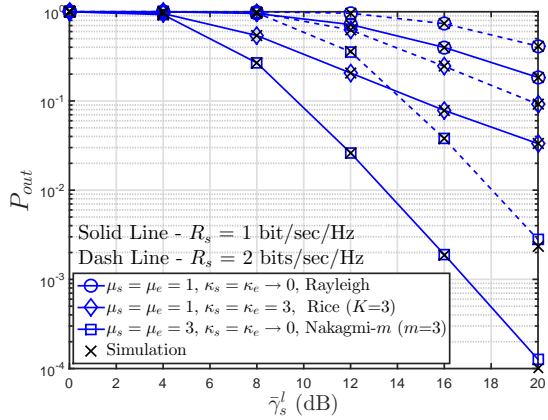}
	\caption{Secrecy Outage Probability against $\bar{\gamma}_S^l$.}
	\label{fig:fig2}
\end{figure}
\begin{align}%
\tau =E_{\gamma _{s}^{l}}\lbrace R_{s}(\gamma _{s}^{l})\rbrace 
=R_{s}\mathcal{P}_{t},%
\end{align}%
where $\mathcal{P}_{t}$ is the transmission probability which can be obtained by using $C^l_s=\textrm{log}_2\left(1+\gamma^l_s\right)$ and \cite[Eq.(3.381)]{gradshteyn2014table} as
\begin{align}%
\mathcal{P}_{t}=1-\prod^L_{l=1}\left[\sum_{w=0}^{\infty }{\frac{(M\mu _{s}\kappa_{s})^{w}\gamma(M\mu _{s}+w,{\Psi_s\mu_{s}(2^{R_{s}}-1)})}{w!\Gamma (M\mu _{s}+w)\mathrm{e}^{(M\mu _{s}\kappa_{s})}}}\right].%
\label{eq_P_t}%
\end{align}%

\section{Results and Discussion}
This section provides numerical results based on the analytical derivations of section 3. Unless stated otherwise, the parameter settings used for the analytical and simulation results shown in this section are as follows: $\bar{\gamma}_S^l$ = 10 dB, $\bar{\gamma}_{ie}^l$=$\bar{\gamma}_{e}^l$ = 0 dB, $PL_s=PL_e=1$ dB, $\rho=0.8$, $\alpha=0.1$ and $\mu_s=\mu_e=1,\ \kappa_s=\kappa_e=1$ $\forall\ i\ \in \ N$. Furthermore, accurate analytical results are achieved by sufficiently considering the first 10 terms of the infinite sum in (\ref{eq_P_l_eve_i_def3}) and (\ref{eq_P_t}).  
\begin{figure}
  \centering
    \begin{tabular}[hb]{c}
      \includegraphics[width=.445\textwidth]{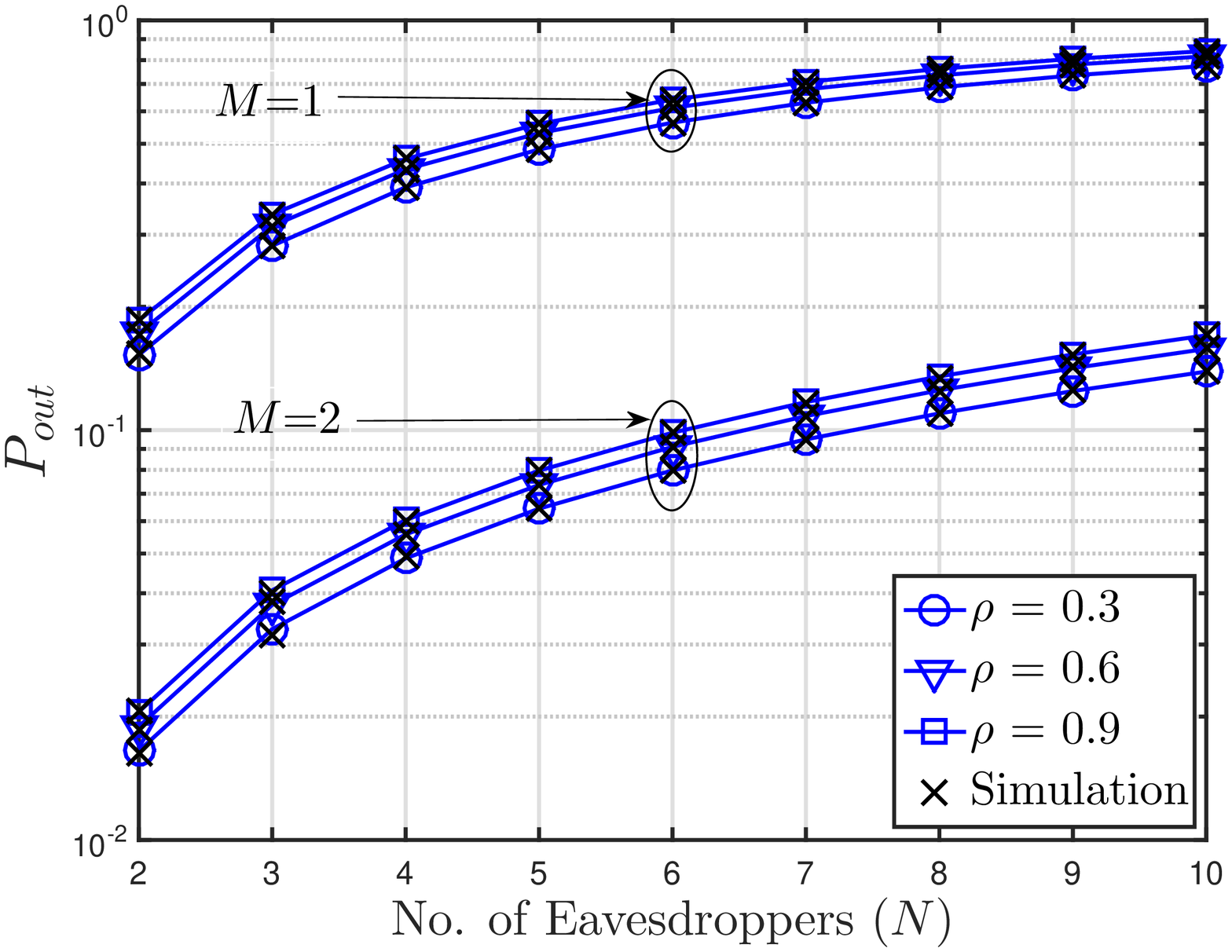}\\
		  \small (a)
	  \end{tabular}\qquad
	\vspace{-0.1em}
	  \begin{tabular}[hb]{c}
		  \includegraphics[width=.445\textwidth]{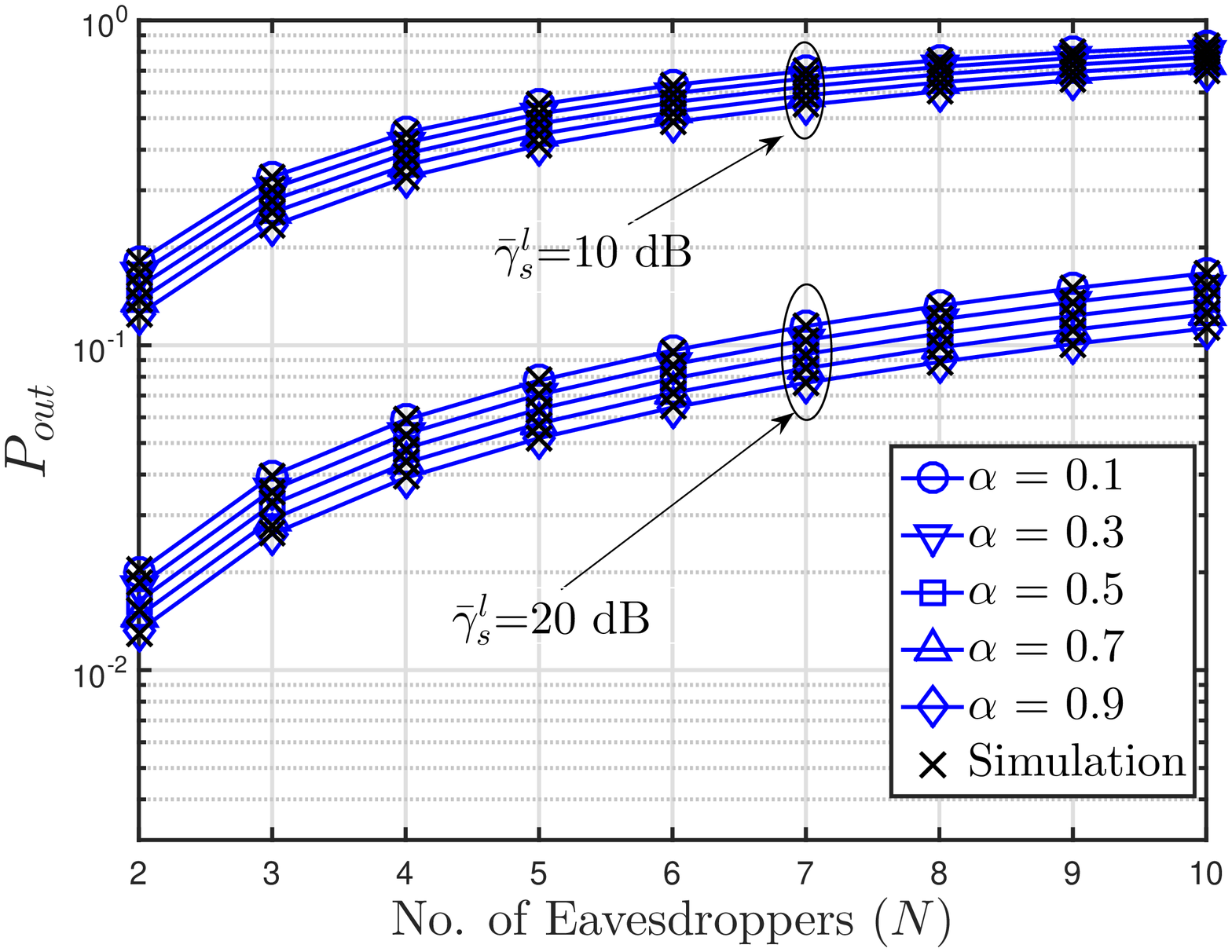}\\
		  \small (b)
	  \end{tabular}%
\caption{Secrecy outage probability versus $N$: (a) effect of $\rho$, (b) effect of $\alpha$.}
\label{fig_3}
\end{figure}
Fig. \ref{fig:fig2} plots the outage probability as a function of $\bar{\gamma}_S^l$ for different channel models. It can be seen that the outage probability increases with the increase in SNR of the main link. Moreover, we observe that Nakagami-$m$ with its parameter $m=3$ outperforms the Rician and Rayleigh fading model which shows that no. of multipath clusters have a significant impact on $P_{out}$. Additionally, the outage probability increases with the increase in $R_s$. However, we observe that the outage performance gap is more in case of Nakagami-$m$ as compared to Rician and Rayleigh for different values of $R_s$.   
\begin{figure}
\centering{\includegraphics[width=.5\textwidth]{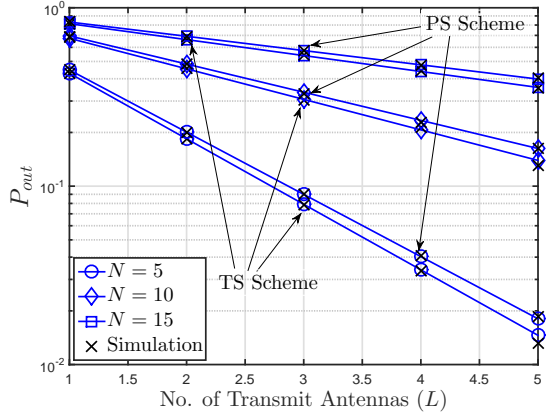}}
\caption{Secrecy outage probability versus $L$ for PS and TS.}
\label{fig_4}
\end{figure}

Fig. \ref{fig_3} shows that the secrecy outage probability increases with increasing $N$ since the cooperative detection ability of eavesdroppers increases with their number. Fig. \ref{fig_3} (a) elaborates on the impact of different values of $\rho$ and $M$ on $P_{out}$; it demonstrates that for the same $N$, an increase in $\rho$ increases $P_{out}$ because larger $\rho$ values indicate that a larger fraction of the received power at eavesdropper is utilized for ID rather than EH. The plot also shows that by increasing $M$, $P_{out}$ decreases. Fig. \ref{fig_3} (b) shows that for a particular $N$, $P_{out}$ decreases with increasing $\alpha$ because larger $\alpha$ values correspond to more fractional time at eavesdroppers spent on EH, during which the ID is off.  

Fig. \ref{fig_4} shows that $P_{out}$ reduces for increasing number of antennas at AP, for both PS and TS at the eavesdroppers; however, the plot shows that the PS architecture achieves a higher secrecy outage probability than that of TS architecture, which indicates more efficient wiretapping performance for the latter. Moreover, it can also be seen from this figure that an increase in $N$ increases the outage probability for both TS and PS.%
\begin{figure}
	\centering
		\includegraphics[width=.5\textwidth]{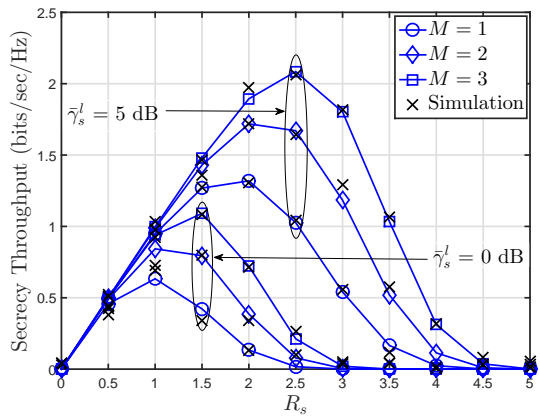}
	\caption{Secrecy Throughput against $R_s$.}
	\label{fig_5}
\end{figure}

Fig. \ref{fig_5} illustrates the impact of increasing $R_s$ on the secrecy throughput; it can be observed from the figure that the secrecy throughput first increases then decreases as $R_s$ is increased. This indicates that there is an optimal value of $R_s$, which maximizes the secrecy throughput. It can also be observed from the figure that increasing $M$ causes an increase in the secrecy throughput, which is the result of increased diversity gain obtained by using more antennas at $S$. Furthermore, the figure shows that for the same $M$, an increase in average SNR of the main link not only increases the secrecy throughput but also shifts its maximal value towards higher values of $R_s$.  
\section{Conclusion}%
We have analyzed physical layer security for the downlink of a SWIPT system in the presence of multiple cooperating eavesdroppers and antenna selection at the AP. A closed-form expression for the secrecy outage probability is derived for $\kappa-\mu$ fading on the links and the impact of power splitting and time-switching parameters of the eavesdroppers on secrecy performance of the considered system have been characterized. Simulation results validate our theoretical analysis.
\section*{Acknowledgments}
This work is supported by the EU-funded project ATOM-690750, approved under call H2020-MSCA-RISE-2015.

\bibliography{mybibfile}
\end{document}